\documentclass[conference,a4paper]{IEEEtran}

\usepackage[latin1]{inputenc}
\usepackage{cite}
\usepackage[pdftex]{graphicx}
 \graphicspath{{../pdf/}{../jpeg/}}
 \DeclareGraphicsExtensions{.pdf,.jpeg,.png}
\usepackage{pgf}
\usepackage{pgfplots}
\usepackage[cmex10]{amsmath}
\usepackage{amssymb}
\usepackage{algorithmic}
\usepackage{array}
\usepackage{mdwmath}
\usepackage{mdwtab}
\usepackage{url}
\usepackage{fixedmultirow}
\usepackage{trfsigns}
\usepackage{fancyhdr}
\usepackage{nicefrac}

\graphicspath{{graphics/}}
\graphicspath{{../Bilder/}}

\usetikzlibrary{calc,trees,positioning,arrows,chains,shapes.geometric,%
 decorations.pathreplacing,decorations.pathmorphing,shapes,%
 matrix,shapes.symbols,plotmarks,decorations.markings,shadows,fit,patterns}
\usetikzlibrary{spy,backgrounds}
\usepgfplotslibrary{groupplots}

\tikzset{every picture/.style={>=latex}} 
\pgfplotsset{filter discard warning=false} 
\pgfplotsset{every axis label/.append style={font=\small}}
\pgfplotsset{every tick label/.append style={font=\footnotesize}}
\pgfsetplotmarksize{2pt}

\tikzset{orientation/.is choice,
    orientation/lr/.style={anchor=west,right=1},
    orientation/lr2/.style={anchor=west,right=2},
    orientation/lrd/.style={anchor=west,below=1},
    orientation/lrd2/.style={anchor=west,below=2},
    orientation/rl/.style={anchor=east,left=1},
    orientation/rl2/.style={anchor=east,left=2},
    orientation/ud/.style={anchor=north,below=1},
    orientation/du/.style={anchor=south,above=1},
    orientation/rld/.style={anchor=east,below=1},
    orientation/rld2/.style={anchor=east,below=2},
}
\tikzstyle{scare} = [
]
\tikzstyle{syslinear} = [
 drop shadow={shadow xshift=.6mm,
              shadow yshift=-.6mm},
 fill=white,
 anchor=west,
 rectangle,
 draw=black,
 minimum height=5mm,
 minimum width=5mm,
 inner xsep=0.5em
]

\tikzstyle{sysnonlinear} = [
 drop shadow={shadow xshift=.8mm,
              shadow yshift=-.8mm},
 fill=white,
 double,
 anchor=west,
 rectangle,
 draw=black,
 minimum height=5mm,
 minimum width=5mm,
 inner xsep=0.5em
]

\tikzstyle{syssource} = [
 anchor=west,
 ellipse,
 draw=black,
 minimum height=1.5ex,
 minimum width=1.5em,
 inner xsep=0.5em
]

\tikzstyle{syssink} = [
 anchor=west,
 ellipse,
 draw=black,
 minimum height=1.5ex,
 minimum width=1.5em,
 inner xsep=0.5em
]

\tikzstyle{syssplit} = [
 fill=black,
 draw=black,
]

\tikzstyle{sysadd} = [
 draw,circle,inner sep=-1pt,
]

\tikzstyle{sysmul} = [
 draw,circle,inner sep=-1pt,
]

\definecolor{MyHSBGreen}{hsb}{0.34065,1,0.91}
\pgfplotscreateplotcyclelist{colors4}{%
 {black!100!white},
 {black!75!white},
 {black!50!white},
 {black!25!white},
}
\pgfplotscreateplotcyclelist{colors5}{%
 {black!100!white},
 {black!75!white},
 {black!50!white},
 {black!25!white},
 {densely dashed, black!100!white},
 {densely dashed, black!75!white},
}
\pgfplotscreateplotcyclelist{colors10}{%
 {black!100!white},
 {black!75!white},
 {black!50!white},
 {black!25!white},
 {densely dashed, black!100!white},
 {densely dashed, black!75!white},
 {densely dashed, black!50!white},
 {densely dashed, black!25!white},
 {densely dotted, black!100!white},
 {densely dotted, black!75!white},
 {densely dotted, black!50!white},
}

\pgfplotscreateplotcyclelist{colorsBERMDDFSE}{%
                 every mark/.append style={fill=black},mark=o\\%
                 every mark/.append style={fill=black},mark=star\\%
                 every mark/.append style={fill=black},mark=triangle\\%
                 every mark/.append style={scale=.7,fill=black},mark=square\\%
                 every mark/.append style={fill=black},mark=diamond\\%
  densely dashed,every mark/.append style={solid},mark=pentagon*\\%
  densely dashed,every mark/.append style={solid},mark=*\\%
  densely dashed,every mark/.append style={solid},mark=triangle*\\%
  densely dashed,every mark/.append style={scale=.7,solid},mark=square*\\%
  densely dashed,every mark/.append style={solid},mark=diamond*\\%
  densely dashed,every mark/.append style={solid},mark=star*\\%
}
\pgfplotscreateplotcyclelist{colors1Empty4Full}{%
                 every mark/.append style={fill=black},mark=o\\%
  densely dashed,every mark/.append style={solid},mark=*\\%
  densely dashed,every mark/.append style={solid},mark=triangle*\\%
  densely dashed,every mark/.append style={scale=.7,solid},mark=square*\\%
  densely dashed,every mark/.append style={solid},mark=diamond*\\%
}

\pgfplotscreateplotcyclelist{colors4Empty4Full}{%
                 every mark/.append style={fill=black},mark=o\\%
                 every mark/.append style={fill=black},mark=triangle\\%
                 every mark/.append style={scale=.7,fill=black},mark=square\\%
                 every mark/.append style={fill=black},mark=diamond\\%
  densely dashed,every mark/.append style={solid},mark=*\\%
  densely dashed,every mark/.append style={solid},mark=triangle*\\%
  densely dashed,every mark/.append style={scale=.7,solid},mark=square*\\%
  densely dashed,every mark/.append style={solid},mark=diamond*\\%
}
\pgfplotscreateplotcyclelist{colors6Empty4Full}{%
                 every mark/.append style={fill=black},mark=o\\%
                 every mark/.append style={fill=black},mark=star\\%
                 every mark/.append style={fill=black},mark=triangle\\%
                 every mark/.append style={scale=.7,fill=black},mark=square\\%
                 every mark/.append style={fill=black},mark=diamond\\%
                 every mark/.append style={fill=black},mark=pentagon\\%
  densely dashed,every mark/.append style={solid},mark=*\\%
  densely dashed,every mark/.append style={solid},mark=triangle*\\%
  densely dashed,every mark/.append style={scale=.7,solid},mark=square*\\%
  densely dashed,every mark/.append style={solid},mark=diamond*\\%
}
\pgfplotscreateplotcyclelist{colors8Empty4Full}{%
                 every mark/.append style={fill=black},mark=o\\%
                 every mark/.append style={fill=black},mark=star\\%
                 every mark/.append style={fill=black},mark=triangle\\%
                 every mark/.append style={scale=.7,fill=black},mark=square\\%
                 every mark/.append style={fill=black},mark=diamond\\%
                 every mark/.append style={fill=black},mark=pentagon\\%
                 every mark/.append style={fill=black},mark=oplus\\%
                 every mark/.append style={fill=black},mark=otimes\\%
  densely dashed,every mark/.append style={solid},mark=*\\%
  densely dashed,every mark/.append style={solid},mark=triangle*\\%
  densely dashed,every mark/.append style={scale=.7,solid},mark=square*\\%
  densely dashed,every mark/.append style={solid},mark=diamond*\\%
}

\pgfplotscreateplotcyclelist{colors}{%
                 every mark/.append style={fill=black},mark=o\\%
                 every mark/.append style={fill=black},mark=star\\%
                 every mark/.append style={fill=black},mark=triangle\\%
                 every mark/.append style={scale=.7,fill=black},mark=square\\%
                 every mark/.append style={fill=black},mark=diamond\\%
                 every mark/.append style={fill=black},mark=pentagon\\%
                 every mark/.append style={fill=black},mark=oplus\\%
                 every mark/.append style={fill=black},mark=otimes\\%
                 every mark/.append style={fill=black},mark=asterisk\\%
                 every mark/.append style={fill=black},mark=+\\%
                 every mark/.append style={fill=black},mark=-\\%
                 every mark/.append style={fill=black},mark=|\\%
                 every mark/.append style={fill=black},mark=x\\%
                 every mark/.append style={fill=black},mark=*\\%
                 every mark/.append style={fill=black},mark=triangle*\\%
                 every mark/.append style={scale=.7,fill=black},mark=square*\\%
                 every mark/.append style={fill=black},mark=diamond*\\%
                 every mark/.append style={fill=black},mark=star*\\%
  densely dashed,every mark/.append style={solid,fill=gray},mark=o\\%
  densely dashed,every mark/.append style={solid,fill=gray},mark=triangle\\%
  densely dashed,every mark/.append style={scale=.7,solid,fill=gray},mark=square\\%
  densely dashed,every mark/.append style={solid,fill=gray},mark=diamond\\%
  densely dashed,every mark/.append style={solid,fill=gray},mark=star\\%
  densely dashed,every mark/.append style={solid,fill=gray},mark=pentagon*\\%
  densely dashed,every mark/.append style={solid,fill=gray},mark=*\\%
  densely dashed,every mark/.append style={solid,fill=gray},mark=triangle*\\%
  densely dashed,every mark/.append style={scale=.7,solid,fill=gray},mark=square*\\%
  densely dashed,every mark/.append style={solid,fill=gray},mark=diamond*\\%
  densely dashed,every mark/.append style={solid,fill=gray},mark=star*\\%
  mark=text,text mark=a\\%
  mark=text,text mark=b\\%
  mark=text,text mark=c\\%
  mark=text,text mark=d\\%
  mark=text,text mark=e\\%
  mark=text,text mark=f\\%
  mark=text,text mark=g\\%
}

\pgfdeclarelayer{background layer}
\pgfdeclarelayer{foreground layer}
\pgfsetlayers{background layer,main,foreground layer}

\newcommand{\ie}{\emph{i.e.}}
\newcommand{\eg}{\emph{e.g.}}
\newcommand{\cf}{\emph{cf.}}

\newcommand{\Eg}{\emph{E.g.}}

\providecommand{\de}[1]{\ensuremath{\mathop{\mathrm{d}}}}

\IEEEoverridecommandlockouts
\title{Punctured Trellis-Coded Modulation}
\author{
 \IEEEauthorblockN{Fabian~Schuh,
                   Andreas~Schenk, and
                   Johannes~B.~Huber}
 \IEEEauthorblockA{Institute for Information Transmission,
                   Friedrich-Alexander-Universit\"at Erlangen-N\"urnberg, Germany\\ 
                   mail: \texttt{\{schuh,\,schenk,\,huber\}@LNT.de}}%
 \thanks{This work was supported by Federal Ministry of Economics and Technology
         (BMWi) within the project C-PMSE.}
}
\tikzstyle{EMTY}       =  [ fill=white, ]
\tikzstyle{STATE}      =  [ fill=black!40!white, ]
\tikzstyle{INPUT}      =  [ fill=black!20!white, ]
\tikzstyle{EXTEND}     =  [ pattern=north east lines, ]
\tikzstyle{TRRSSE}     =  [ pattern=north west lines, ]
\tikzstyle{XSB}        =  [ thick, |-|, shorten <=3pt, shorten >=3pt ]

\begin{document}
\sloppy
\maketitle
\begin{abstract}
THIS PAPER IS ELIGIBLE FOR THE STUDENT PAPER AWARD

In classic trellis-coded modulation (TCM) signal constellations of twice the
cardinality are applied when compared to an uncoded transmission enabling
transmission of one bit of redundancy per PAM-symbol, \ie, rates of
$\frac{K}{K+1}$ when $2^{K+1}$ denotes the cardinality of the signal
constellation. In order to support different rates, multi-dimensional (\ie,
$\mathcal{D}$-dimensional) constellations had been proposed by means of
combining subsequent one- or two-dimensional modulation steps, resulting in
TCM-schemes with $\frac{1}{\mathcal{D}}$ bit redundancy per real dimension.
In contrast, in this paper we propose to perform rate adjustment for TCM by
means of puncturing the convolutional code (CC) on which a TCM-scheme is based
on. It is shown, that due to the nontrivial mapping of the output symbols of
the CC to signal points in the case of puncturing, a modification of the
corresponding Viterbi-decoder algorithm and an optimization of the CC and the
puncturing scheme are necessary.
\end{abstract}
\IEEEpeerreviewmaketitle
\section{Introduction}                                   Ungerboeck's trellis-coded modulation (TCM)~\cite{1056454,UngerboeckTCM87} is
an attractive digital transmission scheme when low over-all delay is desired.
Low latency is ensured by the use of convolutional codes instead of block codes
(\cf~\cite{LIT_tr_com_2009_hehn}) and the dispense with interleaving (as
opposed to convolutional bit-interleaved coded modulation~\cite{141453}).

Ungerboeck showed that a significant increase in Asymptotic Coding Gain (ACG)
can be achieved when considering channel coding and modulation jointly. By
expanding the constellation from $2^{K}$ to $2^{K+1}$ signal points and
employing a rate-$\frac{K}{K+1}$ convolutional encoder one can improve the
noise robustness of the transmission by upto $6\,$dB without any further costs
besides computational effort~\cite{1056454}. High transmission rates can be
achieved when additional uncoded bits are used to address the signal points.
The most important modification to Ungerboeck's TCM propbably is the extension
to multiple dimensions~\cite{1057329} accommodating non-integer transmission
rates by grouping consecutive modulation symbols and constructing a
$\mathcal{D}$-dimensional signal space.

Alternatively, in this paper, puncturing the channel code from a mother code is
considered. As already described in~\cite{31452,502012}, in this case metric
computations become time-dependent, as does the trellis diagram. In contrast to
the modified metric increment suggested in~\cite{502012}, we propose
Maximum-Likelihood (ML) decoding using a modified Viterbi algorithm (VA). Thus
our approach enables ML-decoding of pragmatic punctured TCM for arbitrary
rates. We performed extensive computer simulations in order to optimization the
CC and the puncturing scheme.

This approach can be extended to ISI-channel scenarios. In this case,
ML-decoding can be performed using the matched decoding scheme
of~\cite{Schu1301:Reduced,Schu1301:Matched}.

This paper is structured as follows: First, we briefly introduce notation and
the system model in consideration in Sec.~\ref{sec:systemmodell}.
Sec.~\ref{sec:tcm} recapitulates TCM and introduces a presentation technique
that enables the implementation of punctured encoding in Sec.~\ref{sec:ptcm}.
A code search has been performed to find the best codes for out purpose in
Sec.~\ref{sec:codeSearch}. Final results and conclusions are given in
Sec.~\ref{sec:results} and Sec.~\ref{sec:conclusion}, respectively.

\section{System Model}                                   \label{sec:systemmodell}
\begin{figure}[ht]\vspace*{-2ex}
 \begin{center}
  \begin{tikzpicture}[>=latex,x=1em,y=4ex,font=\footnotesize,inner sep=0.3em,
                      node distance=10mm and 4mm]
   \node[anchor=east] (in) {$u_{\mathrm{c}}[\nu]$};
   \node[syslinear,xshift=5mm,anchor=west,at=(in.east)] (Encoder) {$\mathcal{C}$};
   \draw[o->] (in) -- (Encoder);
   \node[syslinear,right=10mm,anchor=south west,at=(Encoder.south west),minimum width=7mm] (Punct)    {$\mathcal{P}$};
   \node[syslinear,right=5mm, anchor=south west,at=(Punct.south east),minimum height=6.4ex](Labeling) {$\mathcal{L}$};
   \node[syslinear,right=7mm, anchor=west,at=(Labeling.east)]                              (Mapper)   {$\mathcal{M}$};
   \draw[->]  ($(Encoder.east)+(0,.8ex)$) -- ($(Punct.west)+(0,.8ex)$);
   \draw[->]  ($(Encoder.east)-(0,.8ex)$) -- ($(Punct.west)-(0,.8ex)$);
   \path      ($(Punct.west)+(-15mm,3.2ex)$) node[left] {$u_{\mathrm{u}_i}[\nu]$} ++ (7.5mm,0) node {\tiny{$\dots$}} ++ (6mm,0) node[coordinate] (dots) {};
   \draw[o->] ($(Punct.west)+(-15mm,2.4ex)$) -- ++(27.0mm,0);
   \draw[o->] ($(Punct.west)+(-15mm,4.0ex)$) -- ++(27.0mm,0);
   \draw[->]  ($(Punct.east)+(0,.8ex)$) -- ++(5mm,0);
   \draw[->]  ($(Punct.east)-(0,.8ex)$) -- ++(5mm,0);
   \draw (Punct.east) ++(2.3mm,0) ellipse [x radius=2pt,y radius=8pt] node[below=1.6ex] {$c[k]$};
   \draw[->]  (Labeling) -- node[pos=.5,above] {$\ell[k]$} (Mapper);
   \draw[very thin] (dots) +(240:3mm) --        ++(60:3mm) node[above left,font=\tiny,inner sep=1pt] {$n_\mathrm{u}$};
   \draw[very thin] (Punct.west) ++(-3mm,0)     +(240:3mm) node[below right,font=\tiny,inner sep=1pt] {$n_\mathrm{c}$} -- ++(60:3mm);
   \draw[very thin] (Encoder.west) ++(-2mm,0)   +(240:3mm) node[below right,font=\tiny,inner sep=1pt] {$k_\mathrm{c}$} -- ++(60:3mm);
   \draw[-o] (Mapper.east) -- ++(5mm,0) node[right] {$m[k]$};
  \end{tikzpicture}\vspace*{-2ex}
 \end{center}
 \caption{System model for punctured trellis-coded modulation (P-TCM).}
 \label{fig:p-tcm:sysmodel}
 \vspace*{-2ex}
\end{figure}
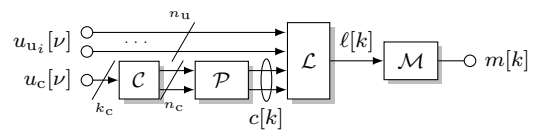

This paper deals with convolutionally encoded pulse-amplitude modulated (PAM)
transmission\footnote{here, the term PAM is used for complex-valued signal
constellations as well including amplitude-shift keying (ASK), phase-shift
keying (PSK) or quadrature-amplitude modulation (QAM)} as depicted in
Fig.~\ref{fig:p-tcm:sysmodel}.
A binary data sequence $\langle u\rangle$ is split into $n_\mathrm{u}$ parallel
uncoded sequences $u_\mathrm{u}[\nu]$ and $k_\mathrm{c}$ parallel sequences
$u_\mathrm{c}[\nu]$ that are encoded using a
rate-$\frac{k_\mathrm{c}}{n_\mathrm{c}}$ binary convolutional $\mathcal{C}$
encoder with generator polynomials $\left[\,g_{ij}\,\right],\; 1 \leq i \leq
n_\mathrm{c}; 1 \leq j \leq k_\mathrm{c}$, with $k_\mathrm{c}$ parallel input
symbols and $n_\mathrm{c}$ parallel output symbols at each time instant.
At each output of the encoder, the symbols traverse through a puncturing system
with puncturing scheme $P = [P_{ij}]\in\left\{ 0,\,1 \right\};\;1<i\leq
n_\mathrm{c};\;1<j<\Omega$ and period $\Omega$. For each encoder input symbol
the puncturing scheme cyclically advances by one step.  Where $P_{ij}$ is zero,
the current symbol at the output is discarded, accordingly.
The punctured $n_\mathrm{c}$-ary encoded output symbols $c[k]$ together with
the uncoded input symbols $u_\mathrm{u}[\nu]$ are labeled to $\ell[k]$ before
being mapped to the $M=2^{n_\mathrm{u}+n_\mathrm{c}}$-ary signal constellation.

Throughout this paper we will use the following notation: 
\begin{itemize}
 \item \mbox{$u[\nu]\in\left\{ 0,\,1 \right\}$}   is the binary input sequence for the
                                       encoder at time instant $\nu$
 \item \mbox{$c[k]\in\left\{ 0,\,1 \right\}$}     denotes the encoded (possibly punctured) symbol
 \item \mbox{$\ell[k]\in\mathcal{L}$}  denotes the signal number from the set of
                                       labelings $\mathcal{L}$ (\eg, natural labeling)
 \item \mbox{$m[k]\in\mathcal{M}$}     is the modulated symbol (\eg, bipolar ASK).
                                       $\mathcal{M}$ denotes the modulation alphabet of
                                       size $M=\left|\mathcal{M}\right|$
 \item \mbox{$P = \left[P_{ij}\right]\in\mathcal{P}$} denotes the puncturing scheme
\end{itemize}

The overall transmission rate of this scheme is $R =
R_\mathrm{c}R_\mathrm{p}n_\mathrm{c} + n_\mathrm{u}$ with
$R_\mathrm{c}=\frac{k_\mathrm{c}}{n_\mathrm{c}}$ and $R_\mathrm{p}$ being the
rate of the channel encoding and puncturing, respectively. The rate of the
puncturing can be calculate from the puncturing scheme by:
\begin{equation}
R_\mathrm{p} = \frac{n_\mathrm{c}\Omega}{\sum\limits_i\sum\limits_j P_{ij}}.
\end{equation}

\section{Trellis-Coded Modulation}                       \label{sec:tcm}

In order to introduce maximum-likelihood sequence estimation (MLSE) for
punctured TCM, we first briefly recall the encoding and mapping process for
\emph{non-punctured} TCM and introduce a representation for the encoding
process and the state transitions of the finite-state-machine (FSM)
(\cf,~\cite{UngerboeckTCM87}).

\subsection{System Model}                                In the case of TCM, the number of output symbols from a
rate-$\frac{k_\mathrm{c}}{n_\mathrm{c}}$ encoder is related to the size of the
modulation alphabet $M$ so that $n_\mathrm{c}=\log_2(M)$ and
$k_\mathrm{c}=n_\mathrm{c}-1$ holds.

\begin{figure}[ht]\vspace{-3ex}
 \begin{center}
  \begin{tikzpicture}[>=latex,x=1em,y=4ex,font=\footnotesize,inner sep=0.3em,
                      node distance=10mm and 4mm]
   \node at (0,0) (u) {$u[k]$};
   \node[syslinear,right=7mm,at=(u.east)] (Encoding) {$\mathcal{C}$};
   \node[syslinear,right=7mm,at=(Encoding.east)] (Labeling) {$\mathcal{L}$};
   \node[syslinear,right=7mm,at=(Labeling.east)] (Mapper) {$\mathcal{M}$};
   \draw[->] ($(Encoding.east)+(0,3pt)$) -- node[midway,above=0.7ex] {$c[k]$} ($(Labeling.west)+(0,3pt)$);
   \draw[->] ($(Encoding.east)-(0,3pt)$) -- ($(Labeling.west)-(0,3pt)$);
   \draw[->] (Labeling) -- node[pos=.5,above=1.3ex] {$\ell[k]$} (Mapper);
   \draw[o->] (u) -- (Encoding);
   \draw[-o] (Mapper.east) -- ++(5mm,0) node[right] {$m[k]$};
   \draw[very thin] (Encoding.east) ++(+3mm,0) +(240:3mm) node[below right,font=\tiny,inner sep=1pt] {$n_\mathrm{c}$} -- ++(60:3mm);
   \draw[very thin] (Encoding.west) ++(-3mm,0) +(240:3mm) node[below right,font=\tiny,inner sep=1pt] {$k_\mathrm{c}$} -- ++(60:3mm);
  \end{tikzpicture}\vspace{-2ex}
 \end{center}
 \caption{Concatenation of a rate-$\frac12$ convolutional encoder
          $\mathcal{C}$, a labeling and modulation. No puncturing is applied
          ($k_\mathrm{u}=0$, $k_\mathrm{c}=1$, $n_\mathrm{c}=2$).}
 \label{fig:SystemModelNonPunct}
 \vspace*{-2ex}
\end{figure}
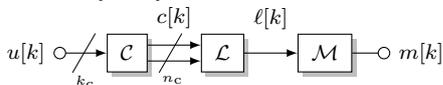

Fig.~\ref{fig:SystemModelNonPunct} shows an example of such a TCM transmitter
with $n_\mathrm{u}=0$, $k_\mathrm{c}=1$ and a rate-$\frac12$ encoder.
Obviously, one of the two channel encoder output symbols (before puncturing)
contains redundancy only.

The coded bits are mapped onto a single transmit symbol $m[k]$ out of a
modulation alphabet of size $2^{n_\mathrm{c}}$. Thus, after labeling and
modulation, the overall transmission rate of TCM is
$R=k_\mathrm{c}+n_\mathrm{u}$. The decoding trellis is well-known,
time-invariant and has $2^{k_\mathrm{c}}$ state transitions for each state.

In the following we describe the encoding process and the FSM to span a
trellis. For simplicity of explanation, we focus on a rate-$\frac12$ encoder
and $4$-ary modulation. We restrict ourselfs to nonrecursive
encoding~\cite{1056454}.

\begin{figure}[ht]
 \begin{center}
  \begin{tikzpicture}[>=latex,x=6em,y=4ex,font=\footnotesize,inner sep=0.3em,
                      node distance=10mm and 4mm]
   \coordinate (u0) at (0,0);
   \coordinate (u1) at (1,0);
   \coordinate (u2) at (2,0);
   \coordinate (u3) at (3,0);
   \draw (u0) circle (1pt) node[above] {$u[k]$};
   \draw (u1) circle (1pt) node[above] {$u[k+1]$};
   \draw (u2) circle (1pt) node[above] {$u[k+2]$};
   \draw (u3) circle (1pt) node[above] {$u[k+3]$};
   \draw (-0.25,-1) circle (1pt) node[coordinate] (c0) {};
   \draw (+0.25,-1) circle (1pt) node[coordinate] (c1) {};
   \draw (+0.75,-1) circle (1pt) node[coordinate] (c2) {};
   \draw (+1.25,-1) circle (1pt) node[coordinate] (c3) {};
   \draw (+1.75,-1) circle (1pt) node[coordinate] (c4) {};
   \draw (+2.25,-1) circle (1pt) node[coordinate] (c5) {};
   \draw (+2.75,-1) circle (1pt) node[coordinate] (c6) {};
   \draw (+3.25,-1) circle (1pt) node[coordinate] (c7) {};
   \begin{scope}[shorten <=1pt,shorten >=1pt]
    \draw[->] (u0) -- node[midway,above,sloped] {$g_1$} (c0);
    \draw[->] (u0) -- node[midway,above,sloped] {$g_2$} (c1);
    \draw[->] (u1) -- node[midway,above,sloped] {$g_1$} (c2);
    \draw[->] (u1) -- node[midway,above,sloped] {$g_2$} (c3);
    \draw[->] (u2) -- node[midway,above,sloped] {$g_1$} (c4);
    \draw[->] (u2) -- node[midway,above,sloped] {$g_2$} (c5);
    \draw[->] (u3) -- node[midway,above,sloped] {$g_1$} (c6);
    \draw[->] (u3) -- node[midway,above,sloped] {$g_2$} (c7);
    \draw (-0.25,-2) node[draw] (s0) {MSB};
    \draw (+0.25,-2) node[draw] (s1) {LSB};
    \draw (+0.75,-2) node[draw] (s2) {MSB};
    \draw (+1.25,-2) node[draw] (s3) {LSB};
    \draw (+1.75,-2) node[draw] (s4) {MSB};
    \draw (+2.25,-2) node[draw] (s5) {LSB};
    \draw (+2.75,-2) node[draw] (s6) {MSB};
    \draw (+3.25,-2) node[draw] (s7) {LSB};
   \end{scope}
   \begin{scope}[shorten <=1pt,shorten >=1pt]
    \foreach \x in {0,...,7} {
     \draw[->] (c\x) -- (s\x);
    }
   \end{scope}
   \draw ($(s0.south west)-(1pt,1pt)$) -| ($(s1.north east)+(1pt,1pt)$) -| ($(s0.south west)-(1pt,1pt)$);
   \draw ($(s2.south west)-(1pt,1pt)$) -| ($(s3.north east)+(1pt,1pt)$) -| ($(s2.south west)-(1pt,1pt)$);
   \draw ($(s4.south west)-(1pt,1pt)$) -| ($(s5.north east)+(1pt,1pt)$) -| ($(s4.south west)-(1pt,1pt)$);
   \draw ($(s6.south west)-(1pt,1pt)$) -| ($(s7.north east)+(1pt,1pt)$) -| ($(s6.south west)-(1pt,1pt)$);
   \begin{scope}[decoration={brace,amplitude=.5em},decorate]
    \draw[decorate] ($(s1.south east)-(0,2pt)$) -- node[yshift=-1ex,midway,below] (l0) {$\ell[k]  $} ($(s0.south west)-(0,2pt)$);
    \draw[decorate] ($(s3.south east)-(0,2pt)$) -- node[yshift=-1ex,midway,below] (l1) {$\ell[k+1]$} ($(s2.south west)-(0,2pt)$);
    \draw[decorate] ($(s5.south east)-(0,2pt)$) -- node[yshift=-1ex,midway,below] (l2) {$\ell[k+2]$} ($(s4.south west)-(0,2pt)$);
    \draw[decorate] ($(s7.south east)-(0,2pt)$) -- node[yshift=-1ex,midway,below] (l3) {$\ell[k+3]$} ($(s6.south west)-(0,2pt)$);
    \draw[->] (l0) -- ++(0,-4ex) node[below] {$m[k]$};
    \draw[->] (l1) -- ++(0,-4ex) node[below] {$m[k+1]$};
    \draw[->] (l2) -- ++(0,-4ex) node[below] {$m[k+2]$};
    \draw[->] (l3) -- ++(0,-4ex) node[below] {$m[k+3]$};
   \end{scope}
  \end{tikzpicture}\vspace*{-2ex}
 \end{center}
 \caption{Encoding process for a rate-$\frac12$ convolutional code and
          $4$-ary natural mapping. Overall transmission rate $R=1$.}
 \label{fig:convEncodingTranditionalProgress}
 \vspace*{-2ex}
\end{figure}
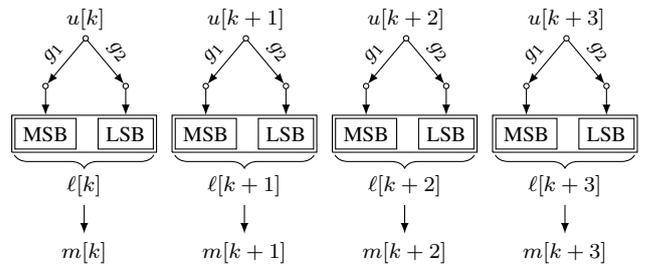

Fig.~\ref{fig:convEncodingTranditionalProgress} illustrates the encoding
process. The uncoded unipolar information sequence $u[k]\in\left\{ 0,\;1
\right\}$ is inserted into the FSM as input values and later passes through all
delay elements describing the state of the FSM. Here, the generator polynomials
$g_1$ and $g_2$ process the input symbol together with the FSM state
synchronously at each time instant. The resulting encoded bits, denoted by
MSB and LSB, respectively, are labeled into $\ell[k]$ and modulated in $m[k]$
to the $4$-ary symbol alphabet of the transmission scheme, \eg, via $m[k] =
2(2\text{MSB}+\text{LSB})-1$ for natural labeling and bipolar ASK (\cf,
Sec~\ref{sec:labeling}).

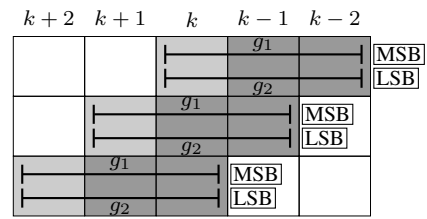
\begin{figure}[ht]\vspace*{-2ex}
 \begin{center}
  \begin{tikzpicture}
   \tikzset{row 1 column 5/.style={nodes={STATE}}}
   \tikzset{row 1 column 4/.style={nodes={STATE}}}
   \tikzset{row 1 column 3/.style={nodes={INPUT}}}
   \tikzset{row 2 column 4/.style={nodes={STATE}}}
   \tikzset{row 2 column 3/.style={nodes={STATE}}}
   \tikzset{row 2 column 2/.style={nodes={INPUT}}}
   \tikzset{row 3 column 3/.style={nodes={STATE}}}
   \tikzset{row 3 column 2/.style={nodes={STATE}}}
   \tikzset{row 3 column 1/.style={nodes={INPUT}}}
   \matrix (FIFO) [matrix of nodes,
                   nodes in empty cells,
                   nodes={draw,
                          ultra thin,
                          anchor=south,
                          rectangle,
                          text width=2em,
                          minimum height=5ex,
                         },
                   ] {
     &&&&\\
     &&&&\\
     &&&&\\
    };
    \draw[XSB] ($(FIFO-1-3.west)+(0,1ex)$) -- node[inner sep=1pt,above,font=\footnotesize,midway] {$g_1$} ($(FIFO-1-5.east)+(0,1ex)$);
    \draw[XSB] ($(FIFO-1-3.west)-(0,1ex)$) -- node[inner sep=1pt,below,font=\footnotesize,midway] {$g_2$} ($(FIFO-1-5.east)-(0,1ex)$);
    \draw[XSB] ($(FIFO-2-2.west)+(0,1ex)$) -- node[inner sep=1pt,above,font=\footnotesize,midway] {$g_1$} ($(FIFO-2-4.east)+(0,1ex)$);
    \draw[XSB] ($(FIFO-2-2.west)-(0,1ex)$) -- node[inner sep=1pt,below,font=\footnotesize,midway] {$g_2$} ($(FIFO-2-4.east)-(0,1ex)$);
    \draw[XSB] ($(FIFO-3-1.west)+(0,1ex)$) -- node[inner sep=1pt,above,font=\footnotesize,midway] {$g_1$} ($(FIFO-3-3.east)+(0,1ex)$);
    \draw[XSB] ($(FIFO-3-1.west)-(0,1ex)$) -- node[inner sep=1pt,below,font=\footnotesize,midway] {$g_2$} ($(FIFO-3-3.east)-(0,1ex)$);
    \node[font=\footnotesize,right=1pt,draw,inner sep=1pt] at ($(FIFO-1-5.east)+(0,1ex)$)  {MSB};
    \node[font=\footnotesize,right=1pt,draw,inner sep=1pt] at ($(FIFO-1-5.east)+(0,-1ex)$) {LSB};
    \node[font=\footnotesize,right=1pt,draw,inner sep=1pt] at ($(FIFO-2-4.east)+(0,1ex)$)  {MSB};
    \node[font=\footnotesize,right=1pt,draw,inner sep=1pt] at ($(FIFO-2-4.east)+(0,-1ex)$) {LSB};
    \node[font=\footnotesize,right=1pt,draw,inner sep=1pt] at ($(FIFO-3-3.east)+(0,1ex)$)  {MSB};
    \node[font=\footnotesize,right=1pt,draw,inner sep=1pt] at ($(FIFO-3-3.east)+(0,-1ex)$) {LSB};
    \node[above,font=\footnotesize] at (FIFO-1-1.north) {$k+2$};
    \node[above,font=\footnotesize] at (FIFO-1-2.north) {$k+1$};
    \node[above,font=\footnotesize] at (FIFO-1-3.north) {$k$};
    \node[above,font=\footnotesize] at (FIFO-1-4.north) {$k-1$};
    \node[above,font=\footnotesize] at (FIFO-1-5.north) {$k-2$};
  \end{tikzpicture}\vspace*{-2ex}
 \end{center}
 \caption{State transitions of the transmitter FSM and the relations between
          generator polynomials and FSM-state/input with $\nu=2$.}
 \label{fig:convEncodingTranditional}
 \vspace*{-2ex}
\end{figure}

Fig.~\ref{fig:convEncodingTranditional} depicts the state transitions of the
FSM, when no puncturing is involved. The states are represented as memory
elements in a first-in first-out (FIFO) register. The time-invariance of the
trellis becomes clear, as each input symbol passes the same encoding procedure
with $g_1$ and $g_2$, and labeling and modulation are independent of the time
instant. As we will see later, this statement does not hold for punctured
coding.
We use the following representation:
\begin{itemize}
 \item dark gray blocks  \mbox{(\!\!\tikz[baseline=0ex]\draw[STATE] rectangle (1.5ex,1.5ex);)} declare the trellis state
 \item light gray blocks \mbox{(\!\!\tikz[baseline=0ex]\draw[INPUT] rectangle (1.5ex,1.5ex);)} define the input values to the FSM and thus, the state transitions
 \item bars              \mbox{(\!\!\tikz[baseline=-.6ex]\draw[XSB] (0,0) -- node[above,inner sep=1pt,font=\footnotesize,midway] {$g_1$} (3em,0);)} describe an arbitrary generator polynomials in binary representation
\end{itemize}

\subsection{Labelings and Modulation}                    \label{sec:labeling}

Labeling and modulation can either be implemented using a lookup table for
$\mathcal{L}$ and $\mathcal{M}$, or via analytical formulas.

\Eg, for the $4$-ary natural labeling we multiply the MSB by $2$ and add the LSB,
\ie, $\ell[k]=2\text{MSB}[k]+\text{LSB}[k]$. An implementation of an ASK
modulator can be seen as unipolar binary symbols $\ell[k]$, mapped onto bipolar
symbols $m[k]$ within an alphabet of size $M$ with $m[k] =
(\ell[k]\cdot2)-(M-1)$.

Different labelings are easily incorporated, \eg, a Gray labeling can be achieved
by $\ell[k] = (1-\text{MSB}[k])(2\text{MSB}[k] + \text{LSB}[k]) + (\text{MSB}[k])
(2 \text{MSB}[k] + (1-\text{LSB}[k]))$.



\subsection{Trellis-based Decoding}                      
\begin{figure}[ht]\vspace{-2ex}
 \begin{center}
  \begin{tikzpicture}[>=latex,x=10em,y=4ex,font=\footnotesize,inner sep=0.3em,
                      node distance=10mm and 10mm]
   \node (u) {$m[k]$};
   \node[coordinate,right=of u,xshift=-12mm] (in) {};
   \draw node[sysadd,right=of in] (pNoise) {$+$};
   \draw node[sysnonlinear, right=of pNoise] (metric) {\parbox{1.5cm}{\centering Metric calculation\\ $\lambda_i[k]$}};
   \draw node[syslinear, right=of metric] (VA) {\parbox{1.5cm}{\centering Trellis-based decoding algorithm}};
   \draw[<-] (pNoise) -- ++(0,6mm) node[above] {$n[k]$};
   \draw[o->] (u.east) -- (pNoise);
   \draw[->] (pNoise) -- (metric);
   \draw[double,double distance=2pt,-implies] (metric) -- (VA);
   \draw[-o] (VA.east) -- ++(6mm,0) node[right] {$\hat{u}[k]$};
  \end{tikzpicture}\vspace{-3ex}
 \end{center}
 \caption{The full-state trellis-based decoding, \eg, with the VA.}
 \label{fig:MDMetricVA}
 \vspace*{-2ex}
\end{figure}
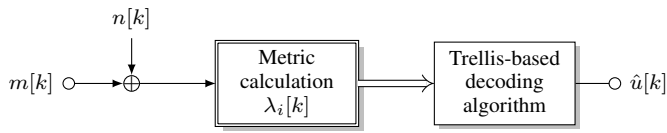

In the following section we will briefly describe the application of a
trellis-based decoding algorithm for Ungerboeck's TCM. The trellis encoder of
the transmitter can be used to generate the hypothesis and the state
transitions of a FSM. The full-state receiver for a non-punctured coding is
depicted in Fig.~\ref{fig:MDMetricVA} and uses the hypothesis to calculate the
metrics $\lambda_i[k]$, \eg, squared Euclidean distances between the received
signal and the hypothesis, for the noisy received signal. The trellis-based
decoding algorithm, such as the VA, estimates the transmitted data sequence by
the use of the trellis and the metrics.

\section{Punctured Trellis-Coded Modulation}             \label{sec:ptcm}
In the following, we will describe the necessary modifications when punctured
coding is applied.

\subsection{System Model}                                
\begin{figure}[ht]\vspace{-2ex}
 \begin{center}
  \begin{tikzpicture}[>=latex,x=1em,y=4ex,font=\footnotesize,inner sep=0.3em,
                      node distance=10mm and 4mm]
   \node at (0,0) (u) {$u[k]$};
   \node[syslinear,right=7mm,at=(u.east)] (Encoding) {$\mathcal{C}$};
   \node[syslinear,right=7mm,at=(Encoding.east)] (Puncturing) {$\mathcal{P}$};
   \node[syslinear,right=7mm,at=(Puncturing.east)] (Labeling) {$\mathcal{L}$};
   \node[syslinear,right=7mm,at=(Labeling.east)] (Mapper) {$\mathcal{M}$};
   \draw[->] ($(Encoding.east)+(0,3pt)$) -- ($(Puncturing.west)+(0,3pt)$);
   \draw[->] ($(Encoding.east)-(0,3pt)$) -- ($(Puncturing.west)-(0,3pt)$);
   \draw[->] ($(Puncturing.east)+(0,3pt)$) -- node[midway,above=0.7ex] {$c[k]$} ($(Labeling.west)+(0,3pt)$);
   \draw[->] ($(Puncturing.east)-(0,3pt)$) -- ($(Labeling.west)-(0,3pt)$);
   \draw[->] (Labeling) -- node[pos=.5,above=1.3ex] {$\ell[k]$} (Mapper);
   \draw[o->] (u) -- (Encoding);
   \draw[-o] (Mapper.east) -- ++(5mm,0) node[right] {$m[k]$};
   \draw[very thin] (Encoding.east) ++(+3mm,0) +(240:3mm) node[below right,font=\tiny,inner sep=1pt] {$n_\mathrm{c}$} -- ++(60:3mm);
   \draw[very thin] (Encoding.west) ++(-3mm,0) +(240:3mm) node[below right,font=\tiny,inner sep=1pt] {$k_\mathrm{c}$} -- ++(60:3mm);
  \end{tikzpicture}\vspace{-2ex}
 \end{center}
 \caption{Concatenation of a rate-$\frac12$ convolutional encoder $\mathcal{C}$
          and puncturing $\mathcal{P}$ with labeling and modulation
          ($k_\mathrm{u}=0$, $k_\mathrm{c}=1$, $n_\mathrm{c}=2$).}
 \label{fig:sysmodel}
 \vspace*{-2ex}
\end{figure}
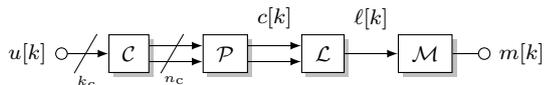

A system model of punctured convolutional coding, without transmission of
uncoded bits, \ie, $n_\mathrm{u}=0$, is depicted in Fig.~\ref{fig:sysmodel}.

By coded modulation with puncturing, we can achieve a transmission rate of $R =
R_\mathrm{c}R_\mathrm{p}R_\mathrm{m}$. For reasons we will see later, our
approach is constraint to $n_\mathrm{c}=2$, \ie, $R_\mathrm{c}=\frac12$. Hence,
the transmission rate is only determined by $R_\mathrm{p} \in [\,1,\,2\,[$.
Note that in the latter, the \emph{mother code} defines the size of the
modulation alphabet while the rate is increased by puncturing.

In the following, we exemplarily focus on $k_\mathrm{u}=0$, $k_\mathrm{c}=1$,
$n_\mathrm{c}=2$ and the puncturing scheme $\left[\,1\,1;\,0\,1\,\right]$.

First, we describe the encoding process and state transitions when
\emph{punctured} codes are used, and discuss the FSM for punctured
convolutional codes as well as the modifications necessary for the VA to run in
the resulting trellis. 

\begin{figure}[ht]
 \begin{center}
  \begin{tikzpicture}
   \tikzset{row 1 column 8/.style={nodes={STATE}}}
   \tikzset{row 1 column 7/.style={nodes={STATE}}}
   \tikzset{row 1 column 6/.style={nodes={INPUT}}}
   \tikzset{row 2 column 7/.style={nodes={STATE}}}
   \tikzset{row 2 column 6/.style={nodes={STATE}}}
   \tikzset{row 2 column 5/.style={nodes={STATE}}}
   \tikzset{row 2 column 4/.style={nodes={INPUT}}}
   \tikzset{row 3 column 6/.style={nodes={STATE}}}
   \tikzset{row 3 column 5/.style={nodes={STATE}}}
   \tikzset{row 3 column 4/.style={nodes={STATE}}}
   \tikzset{row 3 column 3/.style={nodes={INPUT}}}
   \tikzset{row 4 column 4/.style={nodes={STATE}}}
   \tikzset{row 4 column 3/.style={nodes={STATE}}}
   \tikzset{row 4 column 2/.style={nodes={INPUT}}}
   \matrix (FIFO) [matrix of nodes,
                   nodes in empty cells,
                   nodes={draw,
                          ultra thin,
                          anchor=south,
                          rectangle,
                          text width=2em,
                          minimum height=5ex,
                         },
                   ] {
     &&&&&&&\\
     &&&&&&&\\
     &&&&&&&\\
     &&&&&&&\\
    };
    \draw[XSB] ($(FIFO-1-6.west)+(0,1ex)$) -- node[inner sep=1pt,above,font=\footnotesize,midway] {$g_1$} ($(FIFO-1-8.east)+(0,1ex)$);
    \draw[XSB] ($(FIFO-1-6.west)-(0,1ex)$) -- node[inner sep=1pt,below,font=\footnotesize,midway] {$g_2$} ($(FIFO-1-8.east)-(0,1ex)$);
    \draw[XSB] ($(FIFO-2-5.west)+(0,1ex)$) -- node[inner sep=1pt,above,font=\footnotesize,midway] {$g_2$} ($(FIFO-2-7.east)+(0,1ex)$);
    \draw[XSB] ($(FIFO-2-4.west)-(0,1ex)$) -- node[inner sep=1pt,below,font=\footnotesize,midway] {$g_1$} ($(FIFO-2-6.east)-(0,1ex)$);
    \draw[XSB] ($(FIFO-3-4.west)+(0,1ex)$) -- node[inner sep=1pt,above,font=\footnotesize,midway] {$g_2$} ($(FIFO-3-6.east)+(0,1ex)$);
    \draw[XSB] ($(FIFO-3-3.west)-(0,1ex)$) -- node[inner sep=1pt,below,font=\footnotesize,midway] {$g_2$} ($(FIFO-3-5.east)-(0,1ex)$);
    \draw[XSB] ($(FIFO-4-2.west)+(0,1ex)$) -- node[inner sep=1pt,above,font=\footnotesize,midway] {$g_1$} ($(FIFO-4-4.east)+(0,1ex)$);
    \draw[XSB] ($(FIFO-4-2.west)-(0,1ex)$) -- node[inner sep=1pt,below,font=\footnotesize,midway] {$g_2$} ($(FIFO-4-4.east)-(0,1ex)$);
    \node[left] at (FIFO-1-1.west) {$\mathcal{T}_0$};
    \node[left] at (FIFO-2-1.west) {$\mathcal{T}_1$};
    \node[left] at (FIFO-3-1.west) {$\mathcal{T}_2$};
    \node[left] at (FIFO-4-1.west) {$\mathcal{T}_0$};
    \node[font=\footnotesize,right=1pt,draw,inner sep=1pt] at ($(FIFO-1-8.east)+(0,1ex)$)  {MSB};
    \node[font=\footnotesize,right=1pt,draw,inner sep=1pt] at ($(FIFO-1-8.east)+(0,-1ex)$) {LSB};
    \node[font=\footnotesize,right=1pt,draw,inner sep=1pt] at ($(FIFO-2-7.east)+(0,1ex)$)  {MSB};
    \node[font=\footnotesize,right=1pt,draw,inner sep=1pt] at ($(FIFO-2-7.east)+(0,-1ex)$) {LSB};
    \node[font=\footnotesize,right=1pt,draw,inner sep=1pt] at ($(FIFO-3-6.east)+(0,1ex)$)  {MSB};
    \node[font=\footnotesize,right=1pt,draw,inner sep=1pt] at ($(FIFO-3-6.east)+(0,-1ex)$) {LSB};
    \node[font=\footnotesize,right=1pt,draw,inner sep=1pt] at ($(FIFO-4-4.east)+(0,1ex)$)  {MSB};
    \node[font=\footnotesize,right=1pt,draw,inner sep=1pt] at ($(FIFO-4-4.east)+(0,-1ex)$) {LSB};
    \node[above,font=\footnotesize] at (FIFO-1-1.north) {$\nu+5$};
    \node[above,font=\footnotesize] at (FIFO-1-2.north) {$\nu+4$};
    \node[above,font=\footnotesize] at (FIFO-1-3.north) {$\nu+3$};
    \node[above,font=\footnotesize] at (FIFO-1-4.north) {$\nu+2$};
    \node[above,font=\footnotesize] at (FIFO-1-5.north) {$\nu+1$};
    \node[above,font=\footnotesize] at (FIFO-1-6.north) {$\nu$};
    \node[above,font=\footnotesize] at (FIFO-1-7.north) {$\nu-1$};
    \node[above,font=\footnotesize] at (FIFO-1-8.north) {$\nu-2$};
  \end{tikzpicture}\vspace*{-2ex}
 \end{center}
 \caption{State transitions of the transmitter FSM and the relations
          between generator polynomials and FSM-state/input.}
 \label{fig:convEncodingPunctured}
 \vspace*{-2ex}
\end{figure}
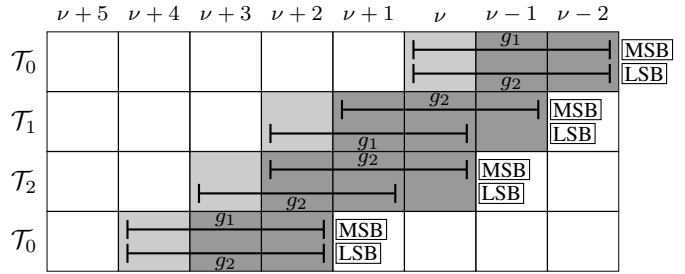

When \emph{punctured} convolutional codes are used, the relation that one
uncoded information bit results in one modulation symbol is not valid any
longer. Note that, whenever the number of erased bits in one period of the
puncturing scheme is not devideable by $\log_2(M)$, the puncturing scheme has
to be repeated until this condition is fulfilled. This restriction ensures that
entire modulation symbols can be constructed by the FSM. In our case, the
puncturing period (\eg, $\left[\,1\,1;\,0\,1\,\right]$) is applied twice. As
can be seen from the encoding procress in
Fig.~\ref{fig:convEncodingPuncturedProgress}, the third and the seventh encoded
symbol is punctured and does not contribute to the labeling and mapping
process. The transitions of the FSM, depicted in
Fig.~\ref{fig:convEncodingPunctured}, show two differences when compared to
\emph{non-punctured} FSM transitions (\cf,
Fig.~\ref{fig:convEncodingTranditional}). These are described subsequently.

\begin{figure}[ht]
 \begin{center}
  \begin{tikzpicture}[>=latex,x=6em,y=4ex,font=\footnotesize,inner sep=0.3em,
                      node distance=10mm and 4mm]
   \coordinate (u0) at (0,0);
   \coordinate (u1) at (1,0);
   \coordinate (u2) at (2,0);
   \coordinate (u3) at (3,0);
   \draw (u0) circle (1pt) node[above] {$u[\nu]$};
   \draw (u1) circle (1pt) node[above] {$u[\nu+1]$};
   \draw (u2) circle (1pt) node[above] {$u[\nu+2]$};
   \draw (u3) circle (1pt) node[above] {$u[\nu+3]$};
   \draw (-0.25,-1) circle (1pt) node[coordinate] (c0) {};
   \draw (+0.25,-1) circle (1pt) node[coordinate] (c1) {};
   \draw (+0.75,-1) circle (1pt) node[coordinate] (c2) {};
   \draw (+1.25,-1) circle (1pt) node[coordinate] (c3) {};
   \draw (+1.75,-1) circle (1pt) node[coordinate] (c4) {};
   \draw (+2.25,-1) circle (1pt) node[coordinate] (c5) {};
   \draw (+2.75,-1) circle (1pt) node[coordinate] (c6) {};
   \draw (+3.25,-1) circle (1pt) node[coordinate] (c7) {};
   \begin{scope}[shorten <=1pt,shorten >=1pt]
    \draw[->] (u0) -- node[midway,above,sloped] {$g_1$} (c0);
    \draw[->] (u0) -- node[midway,above,sloped] {$g_2$} (c1);
    \draw[->] (u1) -- node[midway,above,sloped] {$g_1$} (c2);
    \draw[->] (u1) -- node[midway,above,sloped] {$g_2$} (c3);
    \draw[->] (u2) -- node[midway,above,sloped] {$g_1$} (c4);
    \draw[->] (u2) -- node[midway,above,sloped] {$g_2$} (c5);
    \draw[->] (u3) -- node[midway,above,sloped] {$g_1$} (c6);
    \draw[->] (u3) -- node[midway,above,sloped] {$g_2$} (c7);
    \draw (-0.25,-2) node[draw] (s0) {MSB};
    \draw (+0.25,-2) node[draw] (s1) {LSB};
    \draw (+0.75,-2) node       (s2) {\mbox{ }};
    \draw (+1.25,-2) node[draw] (s3) {MSB};
    \draw (+1.75,-2) node[draw] (s4) {LSB};
    \draw (+2.25,-2) node[draw] (s5) {MSB};
    \draw (+2.75,-2) node       (s6) {\mbox{ }};
    \draw (+3.25,-2) node[draw] (s7) {LSB};
   \end{scope}
   \node at (c2) {\Large\textbf{$\times$}};
   \node at (c6) {\Large\textbf{$\times$}};
   \begin{scope}[shorten <=1pt,shorten >=1pt]
    \foreach \x in {0,1,3,4,5,7} {
     \draw[->] (c\x) -- (s\x);
    }
   \end{scope}
   \draw ($(s0.south west)-(1pt,1pt)$) -| ($(s1.north east)+(1pt,1pt)$) -| ($(s0.south west)-(1pt,1pt)$);
   \draw ($(s3.south west)-(1pt,1pt)$) -| ($(s4.north east)+(1pt,1pt)$) -| ($(s3.south west)-(1pt,1pt)$);
   \draw ($(s5.south west)-(1pt,1pt)$) -| ($(s7.north east)+(1pt,1pt)$) -| ($(s5.south west)-(1pt,1pt)$);
   \begin{scope}[decoration={brace,amplitude=.5em},decorate]
    \draw[decorate] ($(s1.south east)-(0,2pt)$) -- node[yshift=-1ex,midway,below] (l0) {$\ell[k]  $} ($(s0.south west)-(0,2pt)$);
    \draw[decorate] ($(s4.south east)-(0,2pt)$) -- node[yshift=-1ex,midway,below] (l1) {$\ell[k+1]$} ($(s3.south west)-(0,2pt)$);
    \draw[decorate] ($(s7.south east)-(0,2pt)$) -- node[yshift=-1ex,midway,below] (l2) {$\ell[k+2]$} ($(s5.south west)-(0,2pt)$);
    \draw[->] (l0) -- ++(0,-4ex) node[below] {$m[k]$};
    \draw[->] (l1) -- ++(0,-4ex) node[below] {$m[k+1]$};
    \draw[->] (l2) -- ++(0,-4ex) node[below] {$m[k+2]$};
   \end{scope}
  \end{tikzpicture}\vspace*{-2ex}
 \end{center}
 \caption{Encoding process for a rate-$\frac23$ punctured convolutional
          code and natural labeling. Overall transmission rate
          $R=\frac43$.}
 \label{fig:convEncodingPuncturedProgress}
 \vspace*{-2ex}
\end{figure}
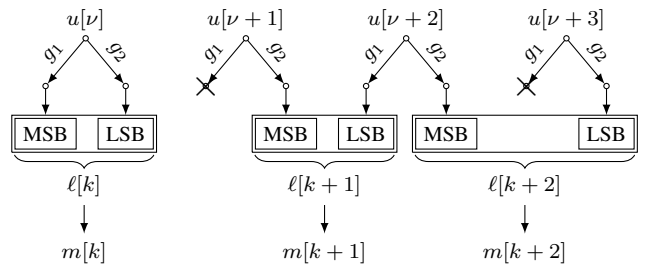

\subsubsection{Generator Offsets $\mathcal{T}_i$}

It becomes clear that the strict relations between the MSB, LSB and the
generator polynomial $g_1$ and $g_2$, respectively, no longer hold. The second
symbol, \ie, $m[\nu+1]$, contains information about $u[\nu+1]$ and $u[\nu+2]$.
In addition, the MSB is now generated by $g_2$ instead of $g_1$.  Accordingly,
the LSB, which was generated by $g_2$ in the \emph{non-punctured} case, is now
generated by $g_1$. It is also clear that the third symbol is generated by
$u[\nu+2]$ and $u[\nu+3]$ using the generator polynomial $g_2$ twice. 

To handle these relations we introduce a set of so called generator offsets
$\mathcal{T}_i$ which describe, depending on the puncturing scheme, modulation
size and time instant, the relations between generator polynomials, input
value, FSM state, and mapping to MSB or LSB, respectively.

In our example, $\mathcal{T}_1$ indicates that the MSB output symbol can be
determined by the memory (\!\tikz[baseline=0ex]\draw[STATE] rectangle
(1.5ex,1.5ex);) only, regardless of the current input symbol.
On the other hand, in order to calculate the LSB we have to use the input value
(\!\tikz[baseline=0ex]\draw[INPUT] rectangle (1.5ex,1.5ex);) and $g_1$.
Obviously, the information in the LSB is one step ahead of the MSB. Thus, we
need to extend the trellis at this point and can calculate the MSB using the
FSM state and the generator polynomial $g_2$.
Note that in case of $M=4$, $\mathcal{T}_0$ indicates that no puncturing is
active, \ie, an even number of puncturings occurred up to time instant $\nu$,
and the generator polynomials are synchronized with LSB and MSB.
$\mathcal{T}_1$ is used, whenever an odd number of puncturings has happened and
$\mathcal{T}_2$ is used when the next puncturing synchronize the generator
polynomials with MSB and LSB again. The number of generator offsets needed to
describe all steps depends on the size of the modulation alphabet, whereas the
generator offsets depend also on the puncturing scheme.

\subsubsection{State Extension}

The FSM transitions in Fig.~\ref{fig:convEncodingPunctured} show that the
symbol $m[\nu+1]$ contains information on the uncoded information bits
$u[\nu+1]$, for which the MSB output was punctured, and the consecutive
information bit $u[\nu+2]$. We see that, when generating the output, the
calculation of the LSB is one step ahead to the MSB and considers one extra
information bit. This results in a trellis that has to be expanded (\ie,
splitted) by a factor of two. This can be easily seen from
Fig.~\ref{fig:convEncodingPunctured} as the generator polynomials in the second
step (generator offsets $\mathcal{T}_1$) now cover four blocks instead of just
three.
However, for a $4$-ary transmission, when puncturing is performed a second
time, MSB and LSB are resynchronized with the generator polynomials and a so
called \emph{merge} happens in the trellis.

When considering $M>4$, a resynchronization will appear after $\log_2(M)$
punctured bits. Extending the FSM to handle $n_\mathrm{c}>2$ does not seem to
be a good alternative choice, due to an increasing number of generator offsets
and the significant increase of computational complexity. Hence, we restrict
ourselves to $n_\mathrm{c}=2$ for complexity reasons.

Therefore, a trellis based decoding algorithm, such as the VA, has to be
performed on a time-variant trellis diagram. An example trellis for a punctured
convolutional code with a constraint length $\nu=2$ will be given below.


\subsection{Trellis-based Decoding}                      \label{sec:decoding}
In the following we will exemplarily describe trellis-based decoding for
punctured convolutional codes.

When using the puncturing scheme $\left[\,1\,0;\,1\,1\,\right]$ introduced
above the VA has to estimate four bits within three symbols from the trellis
transitions. In order to see the modifications of the VA we need to consider
the state extension.

Using Fig.~\ref{fig:convEncodingPunctured}, we define in each step
$\mathcal{T}_0,\dots,\mathcal{T}_2$ the input value to be the first value which
is used by either $g_1$ or $g_2$. As a result, in step $\mathcal{T}_0$ the
values at time instants $k=\left\{ \nu;\;\nu+4 \right\}$ can be defined as
input values.
When in step $\mathcal{T}_1$ or $\mathcal{T}_2$, the input value is at time
instant $k=\left\{ \nu+2;\; \nu+3\right\}$. Unfortunately we cannot estimate
the value for $u[\nu+2]$ in $\mathcal{T}_1$, nor can we estimate $u[\nu+3]$ in
$\mathcal{T}_2$, because one half of the information has not been received,
yet, (\ie, missing information is located in the MSB in $\mathcal{T}_2$ and
$\mathcal{T}_0$, respectively).

However, as all information on $u[\nu+1]$ (for which the output of $g_1$ was
punctured) has been received in $\mathcal{T}_1$ we can estimate it. To do so we
have to evaluate the MSB of the survivor state which is selected by the
survivor path during the add-compare-select procedure within the VA.
In $\mathcal{T}_2$, as already mentioned, selecting the survivor path allows a
decision for two information bits (the MSB of the survivor \emph{and} the input
value) namely $u[\nu+3]$ and $u[\nu+4]$. At this point one has to evalutate the
survivor state for the first information bit, and the survivor transition for
the second one.

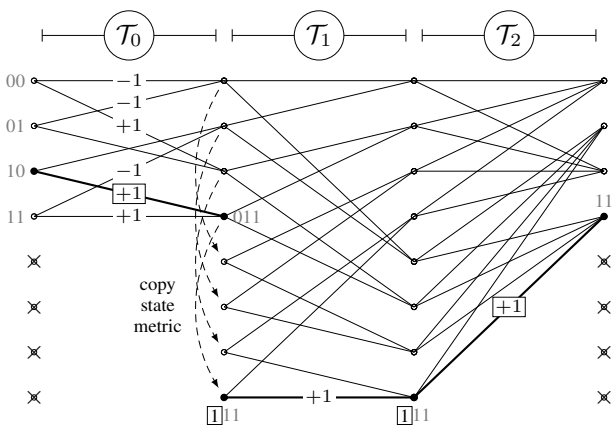
\begin{figure}[ht!]\vspace*{-1ex}
 \begin{center}
  \begin{tikzpicture}[x=25mm,y=6mm,line width=.1pt]
   \draw (1,0) circle(1pt) node {\textbf{$\times$}};
   \draw (1,1) circle(1pt) node {\textbf{$\times$}};
   \draw (1,2) circle(1pt) node {\textbf{$\times$}};
   \draw (1,3) circle(1pt) node {\textbf{$\times$}};
   \draw (4,0) circle(1pt) node {\textbf{$\times$}};
   \draw (4,1) circle(1pt) node {\textbf{$\times$}};
   \draw (4,2) circle(1pt) node {\textbf{$\times$}};
   \draw (4,3) circle(1pt) node {\textbf{$\times$}};
   \draw ( 1,7) circle ( 1pt) -- node[midway,fill=white,inner sep=1pt,font=\scriptsize] {$-1$} ( 2,7) circle ( 1pt);
   \draw ( 1,6) circle ( 1pt) -- node[midway,fill=white,inner sep=1pt,font=\scriptsize] {$-1$} ( 2,7) circle ( 1pt);
   \draw ( 1,5) circle ( 1pt) -- ( 2,6) circle ( 1pt); %
   \draw ( 1,4) circle ( 1pt) -- node[midway,fill=white,inner sep=1pt,font=\scriptsize] {$-1$} ( 2,6) circle ( 1pt);
   \draw ( 1,7) circle ( 1pt) -- node[midway,fill=white,inner sep=1pt,font=\scriptsize] {$+1$} ( 2,5) circle ( 1pt);
   \draw ( 1,6) circle ( 1pt) -- ( 2,5) circle ( 1pt);
   \draw[fill,thick] ( 1,5) circle ( 1pt) -- ( 2,4) circle ( 1pt);
   \path ( 1,5) -- node[draw,midway,fill=white,inner sep=1pt,font=\scriptsize] {$+{1}$} ( 2,4);
   \draw ( 1,4) circle ( 1pt) -- node[midway,fill=white,inner sep=1pt,font=\scriptsize] {$+1$} ( 2,4) circle ( 1pt);
   \node[left,font=\scriptsize,text=gray] at (1,4) {11};
   \node[left,font=\scriptsize,text=gray] at (1,5) {10};
   \node[left,font=\scriptsize,text=gray] at (1,6) {01};
   \node[left,font=\scriptsize,text=gray] at (1,7) {00};
   \draw ( 2,7) circle ( 1pt) -- ( 3,7) circle ( 1pt);
   \draw ( 2,6) circle ( 1pt) -- ( 3,7) circle ( 1pt);
   \draw ( 2,5) circle ( 1pt) -- ( 3,6) circle ( 1pt);
   \draw ( 2,4) circle ( 1pt) -- ( 3,6) circle ( 1pt);
   \draw ( 2,3) circle ( 1pt) -- ( 3,5) circle ( 1pt);
   \draw ( 2,2) circle ( 1pt) -- ( 3,5) circle ( 1pt);
   \draw ( 2,1) circle ( 1pt) -- ( 3,4) circle ( 1pt);
   \draw ( 2,0) circle ( 1pt) -- ( 3,4) circle ( 1pt);
   \draw ( 2,7) circle ( 1pt) -- ( 3,3) circle ( 1pt);
   \draw ( 2,6) circle ( 1pt) -- ( 3,3) circle ( 1pt);
   \draw ( 2,5) circle ( 1pt) -- ( 3,2) circle ( 1pt);
   \draw ( 2,4) circle ( 1pt) -- ( 3,2) circle ( 1pt);
   \draw ( 2,3) circle ( 1pt) -- ( 3,1) circle ( 1pt);
   \draw ( 2,2) circle ( 1pt) -- ( 3,1) circle ( 1pt);
   \draw ( 2,1) circle ( 1pt) -- ( 3,0) circle ( 1pt);
   \draw[fill,thick] ( 2,0) circle ( 1pt) -- ( 3,0) circle ( 1pt);
   \path (2,0) -- node[midway,fill=white,inner sep=1pt,font=\scriptsize] {$+1$} (3,0);
   \node[right,font=\scriptsize,text=gray] at (2,4) {011};
   \node[below,font=\scriptsize,text=gray] at (2,0) {\tikz[baseline=-1.8ex]\node[inner sep=1pt,draw=black,line width=.1pt,text=black]{1};11};
   \draw ( 3,7) circle ( 1pt) -- ( 4,7) circle ( 1pt);
   \draw ( 3,6) circle ( 1pt) -- ( 4,7) circle ( 1pt);
   \draw ( 3,5) circle ( 1pt) -- ( 4,7) circle ( 1pt);
   \draw ( 3,4) circle ( 1pt) -- ( 4,7) circle ( 1pt);
   \draw ( 3,3) circle ( 1pt) -- ( 4,6) circle ( 1pt);
   \draw ( 3,2) circle ( 1pt) -- ( 4,6) circle ( 1pt);
   \draw ( 3,1) circle ( 1pt) -- ( 4,6) circle ( 1pt);
   \draw ( 3,0) circle ( 1pt) -- ( 4,6) circle ( 1pt);
   \draw ( 3,7) circle ( 1pt) -- ( 4,5) circle ( 1pt);
   \draw ( 3,6) circle ( 1pt) -- ( 4,5) circle ( 1pt);
   \draw ( 3,5) circle ( 1pt) -- ( 4,5) circle ( 1pt);
   \draw ( 3,4) circle ( 1pt) -- ( 4,5) circle ( 1pt);
   \draw ( 3,3) circle ( 1pt) -- ( 4,4) circle ( 1pt);
   \draw ( 3,2) circle ( 1pt) -- ( 4,4) circle ( 1pt);
   \draw ( 3,1) circle ( 1pt) -- ( 4,4) circle ( 1pt);
   \draw[fill,thick] ( 3,0) circle ( 1pt) -- ( 4,4) circle ( 1pt);
   \path ( 3,0) -- node[draw,midway,fill=white,inner sep=1pt,font=\scriptsize] {$+{1}$} ( 4,4);
   \node[below,font=\scriptsize,text=gray] at (3,0) {\tikz[baseline=-1.8ex]\node[inner sep=1pt,draw=black,line width=.1pt,text=black]{1};11};
   \node[above,font=\scriptsize,text=gray] at (4,4) {11};
   \draw[shorten <=3pt,shorten >=3pt,|-|] (1,8) -- node[midway,circle,inner sep=2pt,fill=white,draw] {$\mathcal{T}_0$} (2,8);
   \draw[shorten <=3pt,shorten >=3pt,|-|] (2,8) -- node[midway,circle,inner sep=2pt,fill=white,draw] {$\mathcal{T}_1$} (3,8);
   \draw[shorten <=3pt,shorten >=3pt,|-|] (3,8) -- node[midway,circle,inner sep=2pt,fill=white,draw] {$\mathcal{T}_2$} (4,8);
   \draw[densely dashed,->,line width=.2pt,shorten <=4pt,shorten >=4pt] (2,5) to[bend right] (2,1);
   \draw[densely dashed,->,line width=.2pt,shorten <=4pt,shorten >=4pt] (2,6) to[bend right] (2,2);
   \draw[densely dashed,->,line width=.2pt,shorten <=4pt,shorten >=4pt] (2,7) to[bend right] (2,3);
   \draw[densely dashed,->,line width=.2pt,shorten <=4pt,shorten >=4pt] (2,4) to[bend right] node[left,font=\scriptsize,midway] {\parbox{8mm}{\centering copy state metric}} (2,0);
  \end{tikzpicture}\vspace*{-2ex}
 \end{center}
 \caption{Sample Trellis for Decoding}
 \label{fig:trellisPunct}
 \vspace*{-2ex}
\end{figure}

Fig.~\ref{fig:trellisPunct} illustrates where the information bits are
located (boxed) by running the VA over one period of the trellis. As one can
see the first two steps $\mathcal{T}_0$ have two transitions arriving at each
state resulting in an estimation of a single bit per state. However, in the
last step $\mathcal{T}_2$ the decision for a survivor gives an estimate on two
bits. The path register of the VA has to consider the fact that now four bits
have been estimated within three received symbols. The last step can be
described as a state merging, where the split is performed in the first step,
\eg\ by copying the state metrics of the first four states into the last four
states.

\subsection{Increased Transmission Rate}                 In order to increase the transmission rate one can extend the size of the
modulation alphabet and transmit $n_\mathrm{u}$ additional uncoded bits per
modulation step, following the set-partitioning principle~\cite{1056454}. This
TCM approach is well-understood and shall now be extended with puncturing as
depicted in Fig.~\ref{fig:p-tcm:sysmodel}.
By this, the total transmission rate is increased to (here, $R_\mathrm{c}\cdot
n_\mathrm{c}=1$)
\begin{align*}
 R_\mathrm{PTCM} = R_\mathrm{p} + n_\mathrm{u} 
\end{align*}
with $n_\mathrm{u}$ and $n_\mathrm{c}$ denoting the number of uncoded and coded
bits per symbol, respectively. In contrast to Ungerboeck's TCM, our proposed
approach allows to adapt to an arbitrary transmission rate.

The resulting trellis is equal to that in Fig.~\ref{fig:trellisPunct} with
$2^{n_\mathrm{u}}$ parallel branches at each state. Each pair of parallel
branches contain one uncoded bit information per trellis segment. However,
three steps in the time-variant trellis contain four coded information bits.
Thus, the total number of bits per trellis period is seven. We see, that the
transmission of the coded and uncoded bits are not synchronized any longer.

\section{Codes and Puncturing Schemes}                   \label{sec:codeSearch}
As there is no analytical method known to the authors to find the optimal
combination of channel (mother) code, puncturing and labeling for a given
modulation scheme and constraint length, jointly, an exhaustive search was
performed.

The search was carried out for convolutional codes with constraint length $5$
(memory-$4$, \ie, $\nu=4$), natural labeling, and ASK modulation. The
signal-to-noise power ratio $\frac{E_\mathrm{b}}{N_0}$ was between $6\,$ and
$12\,$dB and a transmission of $10^5$ informations bits was performed.

In order to readuce the seach space, the puncturing schemes $P =
\left[P_{ij}\right]\in\mathcal{P};\,i\in\left[ 1,\,2 \right];\,j\in\left[
2,\cdots, \Omega \right]$, with the puncturing period $\Omega$, are used to
define the transmission rate and are constructed in such a way that the sum of
the columns $i$ follow the rule
\begin{align}
 \sum\limits_{i=1}^{2} P_{ij}= 
 \begin{cases}
  2 & \text{if  } j=1\\
  1 & \text{else}.
 \end{cases}
\end{align}
Hence, the rate of the puncturing scheme is $R_\mathrm{p} = \frac{\Omega+1}{\Omega}$.

\begin{table}
 \caption{Best memory-$4$ ($\nu=4$) codes and puncturing schemes for punctured
          convolutional coding with $R_\mathrm{c}=\frac12$, $R_\mathrm{p} =
          \frac{\Omega+1}{\Omega}$.}
 \label{tab:codes}
 \centering
 \begin{tabular}{|c|c|c|l|}\hline
  $R$    & $R_\mathrm{c}\cdot R_\mathrm{p}$ & polynomials & puncturing scheme                                                                      \\\hline\hline
  $4/3$  & $2/3$                      & $\begin{bmatrix}26,\,37\end{bmatrix}_8$ & $\begin{bmatrix}1\,0\\1\,1                        \end{bmatrix}$ \\\hline
  $3/2$  & $3/4$                      & $\begin{bmatrix}36,\,23\end{bmatrix}_8$ & $\begin{bmatrix}1\,1\,1\\1\,0\,0                  \end{bmatrix}$ \\\hline
  $8/5$  & $4/5$                      & $\begin{bmatrix}34,\,31\end{bmatrix}_8$ & $\begin{bmatrix}1\,0\,1 0\\1\,1\,0\,1             \end{bmatrix}$ \\\hline
  $5/3$  & $5/6$                      & $\begin{bmatrix}04,\,37\end{bmatrix}_8$ & $\begin{bmatrix}1\,1\,1\,1\,0\\1\,0\,0\,0\,1      \end{bmatrix}$ \\\hline
  $12/$7 & $6/7$                      & $\begin{bmatrix}34,\,31\end{bmatrix}_8$ & $\begin{bmatrix}1\,0\,1\,0\,1\,0\\1\,1\,0\,1\,0\,1\end{bmatrix}$ \\\hline
 \end{tabular}
\end{table}

The resulting generator polynomials in octal notation and puncturing schemes
for the $16$-state convolutional encoder are listed in Tab.~\ref{tab:codes}.

\section{Numerical Results}                              \label{sec:results}
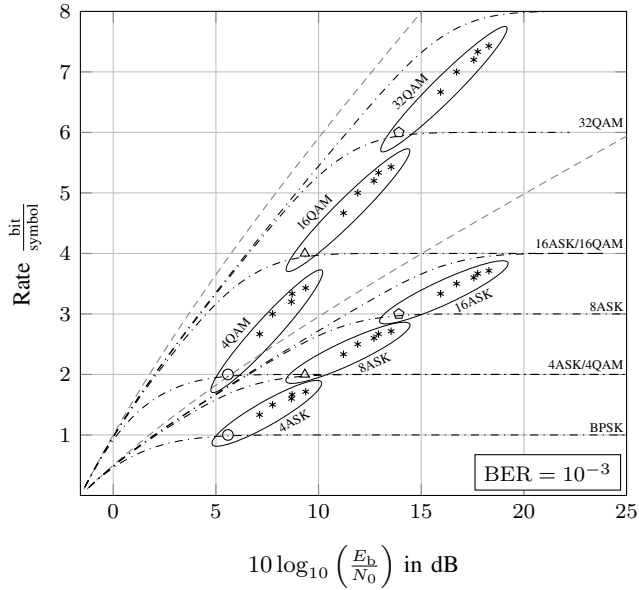
\begin{figure}\vspace*{-2ex}
 \begin{center}
  \begin{tikzpicture}
   \begin{axis}[
                width=.99\linewidth,
                height=8cm,
                xlabel={$10\log_{10}\left(\frac{E_\mathrm{b}}{N_0}\right)$ in dB},
                ylabel={Rate $\frac{\textrm{bit}}{\textrm{symbol}}$},
                cycle list={
                 every mark/.append style={scale=.7,fill=black},mark=asterisk\\%
                 every mark/.append style={scale=.7,fill=black},mark=asterisk\\%
                 every mark/.append style={scale=.7,fill=black},mark=asterisk\\%
                 every mark/.append style={scale=.7,fill=black},mark=asterisk\\%
                 every mark/.append style={scale=.7,fill=black},mark=asterisk\\%
                },
                grid=both,
                xmin=-1.6,xmax=25,
                ymin=0,ymax=8,
                ytick={1,...,8},
                every axis legend/.append style={
                            font={\tiny},
                            nodes={right, inner sep=.1pt}},
                ylabel near ticks,
               ]
    \addplot coordinates {(7.1289910e+00,1.3333333e+00)};
    \addplot coordinates {(7.7586069e+00,1.5000000e+00)};
    \addplot coordinates {(8.6795209e+00,1.6000000e+00)};
    \addplot coordinates {(8.7124434e+00,1.6666667e+00)};
    \addplot coordinates {(9.3708797e+00,1.7142857e+00)};
    \addplot coordinates {(1.1215127e+01,2.3333333e+00)};
    \addplot coordinates {(1.1906486e+01,2.5000000e+00)};
    \addplot coordinates {(1.2697113e+01,2.6000000e+00)};
    \addplot coordinates {(1.2917490e+01,2.6666667e+00)};
    \addplot coordinates {(1.3531312e+01,2.7142857e+00)};
    \addplot coordinates {(1.5942384e+01,3.3333333e+00)};
    \addplot coordinates {(1.6721482e+01,3.5000000e+00)};
    \addplot coordinates {(1.7560098e+01,3.6000000e+00)};
    \addplot coordinates {(1.7750679e+01,3.6666667e+00)};
    \addplot coordinates {(1.8295241e+01,3.7142857e+00)};
    \addplot[mark=o]        coordinates {(5.5855856e+00,1.0000000e+00)};
    \addplot[mark=triangle] coordinates {(9.3320424e+00,2.0000000e+00)};
    \addplot[mark=pentagon] coordinates {(1.3907466e+01,3.0000000e+00)};
    \coordinate (A) at (axis cs:5.5855856e+00,1.0000000e+00);
    \coordinate (B) at (axis cs:9.3708797e+00,1.7142857e+00);
    \node[line width=.1pt,inner sep=0pt,minimum height=2ex,draw,rotate fit=30,ellipse,fit=(A) (B),label={[inner sep=1pt,font=\tiny,rotate=30]below:4ASK}] {};
    \coordinate (A) at (axis cs:9.3320424e+00,2.0000000e+00);                                                                    
    \coordinate (B) at (axis cs:1.3531312e+01,2.7142857e+00);                                                                    
    \node[line width=.1pt,inner sep=0pt,minimum height=2ex,draw,rotate fit=25,ellipse,fit=(A) (B),label={[inner sep=1pt,font=\tiny,rotate=25]below:8ASK}] {};
    \coordinate (A) at (axis cs:1.3907466e+01,3.0000000e+00);                                                                    
    \coordinate (B) at (axis cs:1.8295241e+01,3.7142857e+00);                                                                    
    \node[line width=.1pt,inner sep=0pt,minimum height=2ex,draw,rotate fit=25,ellipse,fit=(A) (B),label={[inner sep=1pt,font=\tiny,rotate=25]below:16ASK}] {};
    \addplot[densely dashed,gray] table[x index=0,y index=1] {data-CapacityShannon.csv};
    \addplot[smooth,black,dashdotted] table[x index=0,y index=1] {data-CmASK.csv};
    \addplot[smooth,black,dashdotted] table[x index=2,y index=3] {data-CmASK.csv};
    \addplot[smooth,black,dashdotted] table[x index=4,y index=5] {data-CmASK.csv};
    \addplot[smooth,black,dashdotted] table[x index=6,y index=7] {data-CmASK.csv};
    \addplot coordinates {(7.1289910e+00,1.3333333e+00*2)};
    \addplot coordinates {(7.7586069e+00,1.5000000e+00*2)};
    \addplot coordinates {(8.6795209e+00,1.6000000e+00*2)};
    \addplot coordinates {(8.7124434e+00,1.6666667e+00*2)};
    \addplot coordinates {(9.3708797e+00,1.7142857e+00*2)};
    \addplot coordinates {(1.1215127e+01,2.3333333e+00*2)};
    \addplot coordinates {(1.1906486e+01,2.5000000e+00*2)};
    \addplot coordinates {(1.2697113e+01,2.6000000e+00*2)};
    \addplot coordinates {(1.2917490e+01,2.6666667e+00*2)};
    \addplot coordinates {(1.3531312e+01,2.7142857e+00*2)};
    \addplot coordinates {(1.5942384e+01,3.3333333e+00*2)};
    \addplot coordinates {(1.6721482e+01,3.5000000e+00*2)};
    \addplot coordinates {(1.7560098e+01,3.6000000e+00*2)};
    \addplot coordinates {(1.7750679e+01,3.6666667e+00*2)};
    \addplot coordinates {(1.8295241e+01,3.7142857e+00*2)};
    \addplot[mark=o]        coordinates {(5.5855856e+00,1.0000000e+00*2)};
    \addplot[mark=triangle] coordinates {(9.3320424e+00,2.0000000e+00*2)};
    \addplot[mark=pentagon] coordinates {(1.3907466e+01,3.0000000e+00*2)};
    \coordinate (A) at (axis cs:5.5855856e+00,2);
    \coordinate (B) at (axis cs:9.3708797e+00,3.4286);
    \node[line width=.1pt,inner sep=0pt,minimum height=2ex,draw,rotate fit=48,ellipse,fit=(A) (B),label={[inner sep=1pt,font=\tiny,rotate=48]above:4QAM}] {};
    \coordinate (A) at (axis cs:9.3320424e+00,4);
    \coordinate (B) at (axis cs:1.3531312e+01,5.4286);
    \node[line width=.1pt,inner sep=0pt,minimum height=2ex,draw,rotate fit=45,ellipse,fit=(A) (B),label={[inner sep=1pt,font=\tiny,rotate=45]above:16QAM}] {};
    \coordinate (A) at (axis cs:1.3907466e+01,6);
    \coordinate (B) at (axis cs:1.8295241e+01,7.4286);
    \node[line width=.1pt,inner sep=0pt,minimum height=2ex,draw,rotate fit=45,ellipse,fit=(A) (B),label={[inner sep=1pt,font=\tiny,rotate=45]above:32QAM}] {};
    \addplot[densely dashed,gray] table[x index=2,y index=3] {data-CapacityShannon.csv};
    \addplot[smooth,black,dashdotted] table[x index=0,y index=1] {data-CmQAM.csv};
    \addplot[smooth,black,dashdotted] table[x index=2,y index=3] {data-CmQAM.csv};
    \addplot[smooth,black,dashdotted] table[x index=4,y index=5] {data-CmQAM.csv};
    \addplot[smooth,black,dashdotted] table[x index=6,y index=7] {data-CmQAM.csv};
    \node[anchor=south east,font=\tiny,inner sep=1pt] at (axis cs:25,1) {BPSK};
    \node[anchor=south east,font=\tiny,inner sep=1pt] at (axis cs:25,2) {4ASK/4QAM};
    \node[anchor=south east,font=\tiny,inner sep=1pt] at (axis cs:25,3) {8ASK};
    \node[anchor=south east,font=\tiny,inner sep=1pt] at (axis cs:25,4) {16ASK/16QAM};
    \node[anchor=south east,font=\tiny,inner sep=1pt] at (axis cs:25,6) {32QAM};
    \node[draw,fill=white,anchor=south east,font={\footnotesize}] at (rel axis cs:0.99,0.01) {$\mathrm{BER} = 10^{-3}$};
   \end{axis}
  \end{tikzpicture}\vspace*{-2ex}
 \end{center}
 \caption{Spectral efficiency vs. required $\frac{E_\mathrm{b}}{N_0}$ for a
          bit error rate of $\mathrm{BER}=10^{-3}$. Channel codes with $2^4$
          states. Shannon capacity constraint (dashed line,
          real-/complex-valued), Constellation capacity (dash-dotted line, ASK and
          QAM). Asterisk markers: proposed approach. Empty markers:
          Ungerboeck's TCM, $\mathcal{D}=1$}
 \label{fig:spectralEfficiency}
 \vspace*{-2ex}
\end{figure}

Fig.~\ref{fig:spectralEfficiency} shows the simulation results for ASK 
modulation schemes. The $16$-state channel codes are punctured from a rate
$\frac12$ mother code using the puncturing scheme as defindes in
Table~\ref{tab:codes}. Higher transmission rates are achieved by additional
uncoded bits and expansion of the constellation (\cf,
Fig.~\ref{fig:p-tcm:sysmodel}), following the set partitioning principle.

Additionally, two consecutive ASK channel symbols are modulated to a
two-dimensional quadrature-amplitude-modulation (QAM) symbol to further
increase transmission rate without increasing the signal bandwidth.

The results clearly indicate that punctured TCM enables soft transitions
between classical TCM rates (integers) by puncturing. By this means, the shape
of the constellation constraint capacity can be closely approximated using
punctured TCM.

\section{Conclusion}                                     \label{sec:conclusion}

It has been shown that TCM can be extended by puncturing. We sketched the
modifications necessary for the VA to properly work on the resulting
time-variant trellises. Additionally, we conducted a computer search to
optimize for the channel encoder and the puncturing scheme for ASK modulation.
We further increased the transmission rate by appending uncoded information
bits and thus enlarging the signal constellation as well as by implementing
QAM transmission scheme using two subsequent ASK symbols.

The numerical simulation results clearly show that we can achieve a soft
tradeoff between spectral and power efficiency easier and more flexible than by
means of traditional TCM and multidimensional TCM.



\bibliographystyle{IEEEtran}
\bibliography{IEEEabrv,main,mine}

\end{document}